# Dynamics of a polariton condensate transistor switch

C. Antón,[1, 2, *] T.C.H. Liew,[3] G. Tosi,[1] M.D. Martín,[1, 2] T. Gao,[4, 5] Z. Hatzopoulos,[5,6] P.S. Eldridge,[5] P.G. Savvidis,[4, 5] and L. Viña[1, 2]

[1]*Departamento de Física de Materiales, Universidad Autónoma de Madrid, 28049 Madrid, Spain*
[2]*Instituto de Ciencia de Materiales "Nicolás Cabrera", Universidad Autónoma de Madrid, 28049 Madrid, Spain*
[3]*School of Physical and Mathematical Sciences, Nanyang Technological University, 637371, Singapore*
[4]*Department of Materials Science and Technology, Univ. of Crete, 71003 Heraklion, Crete, Greece*
[5]*FORTH-IESL, P.O. Box 1385, 71110 Heraklion, Crete, Greece*
[6]*Department of Physics, University of Crete, 71003 Heraklion, Crete, Greece*

We present a time-resolved study of the logical operation of a polariton condensate transistor switch. Creating a polariton condensate (source) in a GaAs ridge-shaped microcavity with a non-resonant pulsed laser beam, the polariton propagation towards a collector, at the ridge edge, is controlled by a second weak pulse (gate), located between the source and the collector. The experimental results are interpreted in the light of simulations based on the generalized Gross-Pitaevskii equation, including incoherent pumping, decay and energy relaxation within the condensate.

All optical devices, as an alternative to electronic ones, due to their larger bandwidth, reduced noise and power dissipation, have been sought during the last two decades for the realization of ultrafast communication and signal processing systems. Semiconductor materials are the preferred choice for the realization of such devices (see Ref. 1). Excitonic-based circuits have been proposed[2] and demonstrated.[3] Exciton-polaritons, due to their mixed light-matter nature and their optical nonlinearities, are even more suitable to create optical amplifiers,[4-6] optical gates,[7,8] transistors,[9,10] spin-switches,[11-15] and circuits.[16] Further functionalities can be achieved creating polariton condensates and reducing the dimensionality by patterning the microcavities. Propagation of such condensates over macroscopic distances has been achieved in wire microcavities with very long polariton lifetimes.[17] The condensates can be conveniently manipulated using repulsive local potentials created by photogeneration of excitons.[18] Large band-width amplification of polariton condensates under non-resonant excitation has been proven by a proper location of the laser excitation spot to create a condensate close to the edge of a 1D microwire.[19] In these systems, thanks to their superfluid character, one can benefit from high lateral speed of propagation and ballistic transport without energy loss.

Using wider microwire ridges, a polariton condensate transistor switch has been realised through optical excitation with two beams.[9] One of the beams creates a polariton condensate which serves as a source (*S*) of polaritons. Their propagation is gated using a second weaker gate beam (*G*) that controls the polariton flow by creating a local blueshifted barrier. The ON state of the transistor (no *G*) corresponds to forming a trapped condensate at the edge of the ridge (collector, *C*). The presence of *G* hinders the propagation of polaritons towards *C*, remaining blocked between *S* and *G* (OFF state). In this letter, we present a time-resolved study that provides a complete insight of the energy relaxation and dynamics of the condensed polariton propagation for the ON/OFF states of such a transistor switch. Our experiments are compared with a theoretical description of the polariton condensate transistor based on the generalized Gross-Pitaevskii equation, modified to account for incoherent pumping, decay and energy relaxation within the condensate.

We investigate a high-quality $5\lambda/2$ AlGaAs-based microcavity with a Rabi splitting $\Omega_R = 9$ meV. Ridges have been sculpted through reactive ion etching with dimensions 20 × 300 µm² (further information about this sample is given in Refs. 20 and 21). In our samples, lateral confinement is insignificant compared to the aforementioned, much thinner, 1D polariton wires.[17,19] The sample, mounted in a cold-finger cryostat and kept at 10 K, is excited with 2 ps-long light pulses from a Ti:Al$_2$O$_3$ laser, tuned to the first high-energy Bragg mode of the distributed Bragg reflector (1.612 eV). The chosen ridge is in a region of the sample corresponding to resonance (detuning between the bare exciton and cavity mode ~ 0 meV). We split the laser beam to produce two twin beams, whose intensities, spatial positions and relative time delay can be independently adjusted. We focus both beams on the sample through a microscope objective to form 5-µm ⌀ spots spatially separated by 40 µm along the ridge. The source spot is placed 80 µm away from the right ridge border. The same objective is used to collect and direct the emission, towards a spectrometer, to a streak camera obtaining energy-, time- and spatial-resolved images, with resolutions of 0.4 meV, 15 ps and 1 µm, respectively.

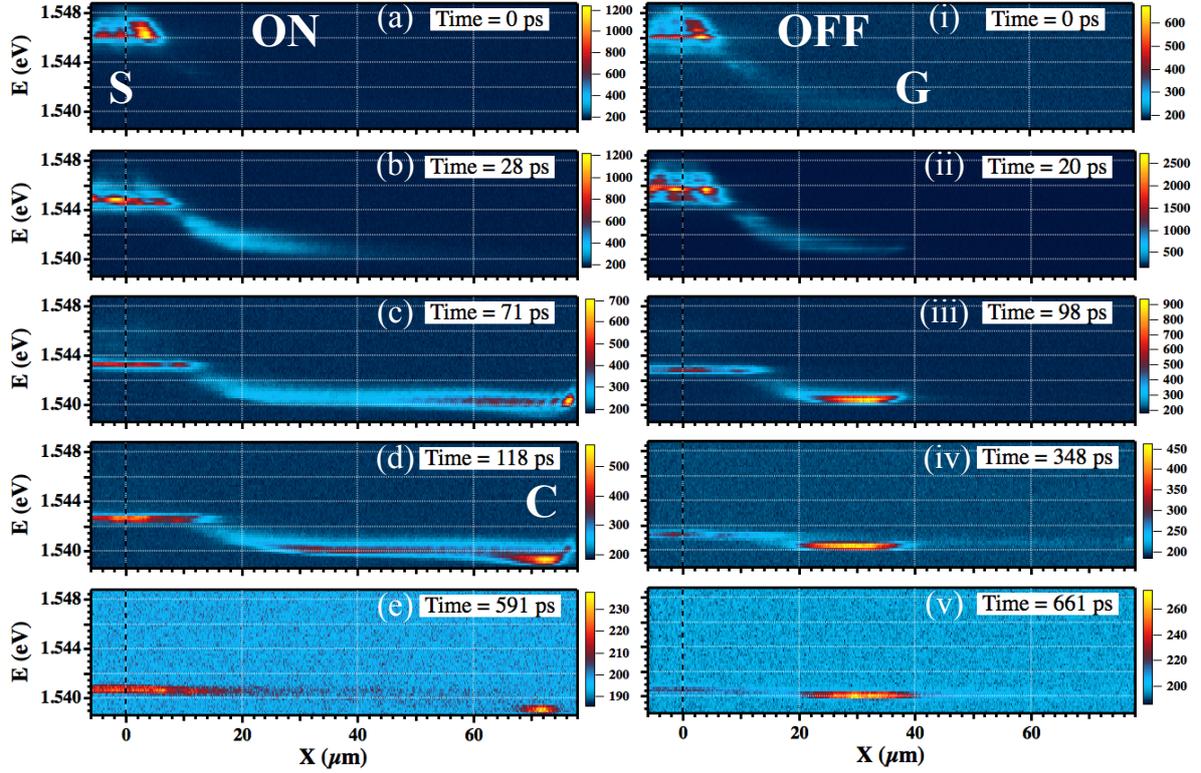

Figure 1. (a-e)/(i-v) Energy vs. position for the ON/OFF transistor state. The snapshots are taken at the times shown in the labels. The intensity is coded in a false color scale shown on the right of each panel.

From the different configurations presented in Fig. 3 of Ref. 9, where the S power is kept constant and the G one is varied, we present here the dynamics of the two most relevant cases, corresponding to the ON (OFF) state of the switch, given by the powers $P_S$= 10.8 mW and $P_G$= 0 (0.6) mW, respectively. $P_S$ correspond to ~7 times the power threshold for condensation of polaritons ($P_{th}$=1.5 mW), while $P_G$ is well below threshold. Figures 1 (a-e) show the dynamics for the ON state. X=0 µm and time 0 ps, in the panels, are chosen at the S position and when the emission reaches its maximum intensity. In panel (a), the initial source signal, at an energy of ~1.546 eV, arises from the emission of large momentum excitons, extending ~10 µm. At longer times, these excitons relax rapidly their energy and concomitantly polaritons that are formed at S expand in space. These two processes give rise to an emission at S relaxing rapidly its energy in a doubly exponential fashion with decay times $\tau_1$= 18 ps and $\tau_2$= 270 ps; the long $\tau_2$ corresponds to the polariton energy relaxation. Simultaneously with these decay processes, two polariton populations emerge at lower energy than that of S; they travel left (not shown) and right away from the excitation spot due to the local energy potential landscape.[17] Initially the right travelling polaritons decrease their energy as they move away from S (b), until a condensate propagates ballistically and reaches C, see (c), the interferences appearing around 65 µm corroborate that the flowing polaritons are in a condensed state. At later times (d), a large fragment of this population becomes trapped just before the end of the ridge, at 1.539 eV, while a smaller part bounces back from the edge and travels back towards *S* at a constant energy ~1.540 eV. At very long times, ~600 ps (e), only the *C*-trapped polaritons and the *S*-polaritons remain. The gradual decrease of the emission intensity is due to recombination processes.

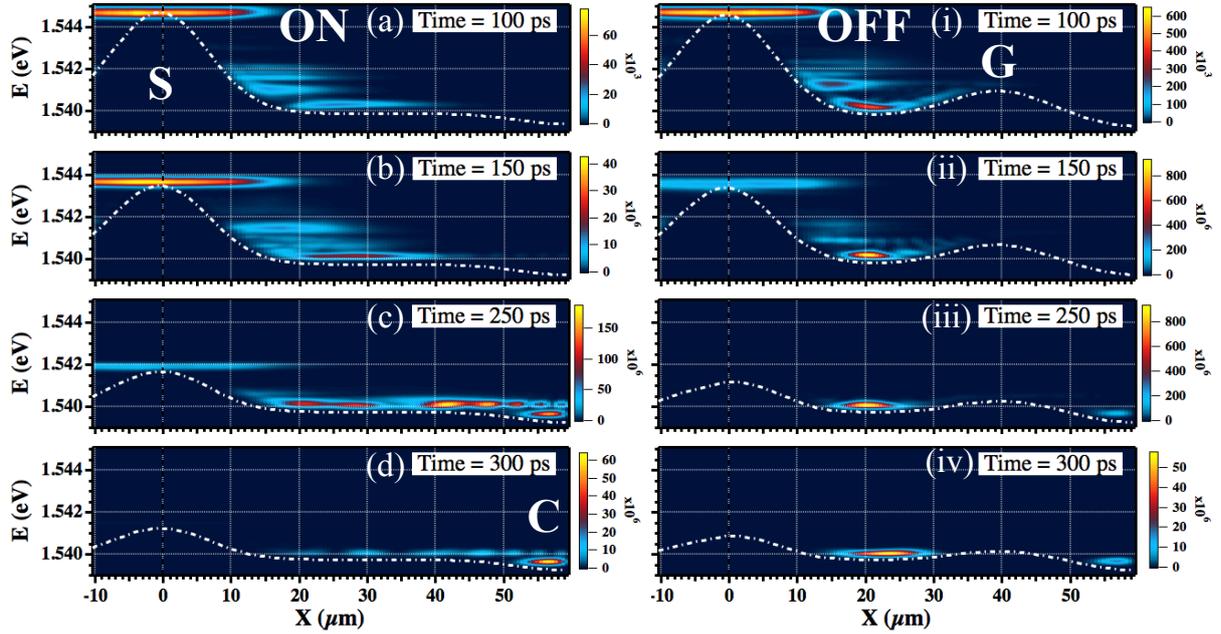

Figure 2. Theoretical dynamics of the energy-position distribution in the ON (a-d)/OFF (i-iv) transistor state, evaluated from Fourier transforms of Ψ(x,t). The white dashed lines show the effective potential experienced by polaritons, V(x,t). On the right of each panel, the false color code of the intensity is depicted in arbitrary units.

The corresponding dynamics of the OFF state is compiled in Figs. 1(i-v). In this case, the *G*-laser beam, which arrives simultaneously with the *S*-beam, is placed at 40 μm and constitutes an invisible barrier as its power is only $0.4P_{th}$.[22] The right propagating polaritons display a similar initial behaviour to that shown before, compare (b) with (ii). However, a very large population, which dominates the emission (iii), gets trapped between *S* and *G*. It emits at a slightly larger energy, 1.541 eV, than that of *C*, due to the addition of the *S* and *G* potential profiles and to its own blueshift. When the *G* barrier becomes lower (iv), a tiny part of the trapped population tunnels through and reaches the border of the ridge without becoming confined at *C*. At long times (v), the emission arises only from gated polaritons. A more detailed dynamics of the ON/OFF state can be found in the supplementary multimedia files (ONstate.mov and OFFstate.mov).

A theoretical description of the polariton condensate transistor can be based on the generalized Gross-Pitaevskii equation for the condensed polariton wavefunction $\psi(x,t)$:[23]

$$i\hbar \frac{d\psi(x,t)}{dt} = [\hat{E}_{LP} + \alpha |\psi(x,t)|^2 + V(x,t) + \frac{i\hbar}{2}(Rn(x,t)-\gamma)]\psi(x,t) + i\hbar\Re[\psi(x,t)], \quad (1)$$

where $\hat{E}_{LP}$ is the polariton kinetic energy operator; $\alpha$ is the polariton-polariton interaction strength; $V(x,t) = V_0(x) + gn(x,t)$ is the polariton potential, given by the static potential defining the wire edge, $V_0(x)$, and a potential shift, proportional to the constant $g$, due to the presence of a hot exciton reservoir, $n(x,t)$; $R$ is the condensation rate; $\gamma$ is the decay rate. The incoming rate of polaritons into the condensate is determined by the exciton reservoir density, which was calculated from the evolution equation given in Ref. 23. This theory, or similar versions, has been used to describe a variety of recent experiments under incoherent pumping, including the dynamics of vortices[24] as well as various spatial patterns[25,26] and spin textures.[27] The last term in Eq. 1 accounts for the energy relaxation of condensed polaritons, which may occur due to polariton-phonon scattering. Experimentally, the energy relaxation of polariton condensates is most clearly observed as they propagate down static[9,28] or pump induced[17] potentials. Previous descriptions of energy relaxation have been based on the introduction of an additional decay of particles depending on their energy[19,29-31] (occasionally known as the Landau-Khalatnikov approach):

$$\Re[\psi(x,t)] = -\nu(\hat{E}_{LP} - \mu(x,t))\psi(x,t) \quad (2)$$

where $\nu$ is a phenomenological parameter determining the strength of energy relaxation[19,29-31] and $\mu(x,t)$ is a local effective chemical potential, which is chosen such that $\Re[\psi(x,t)]$ does not change the number of particles at each point in space.[31] As in Ref. 19, we assume the energy relaxation rate to be proportional to the kinetic energy of polaritons; polaritons relax in energy until they decay from the system or until their kinetic energy is zero.

The calculated evolution of the energy-position distribution[32] is shown in Fig. 2 and reproduces the experimental findings. As polaritons move away from S, their potential energy decreases and is converted into kinetic energy. Due to the term in Eq. 2, this kinetic energy is relaxed and the polariton energy follows the effective potential, $V(x,t)$. In the absence of G (ON state), the polaritons can be reflected from the ridge edge giving rise to the interference fringes seen at t=250 ps (Fig. 2(c)). At longer times, further energy relaxation results in trapping in a potential minimum near the ridge edge (C), whose depth was inferred from the experimental results. In addition, the fraction of polaritons that remain at the source position also relax their energy, as the effective potential lowers due to decay of the exciton reservoir.

The blocking of polaritons in the presence of G (OFF state) is also observed, in agreement with the experiments, where polaritons are trapped between S and G. Some quantum tunnelling across G occurs, particularly at later times when the barrier height has decreased.

While the theoretical model captures the main qualitative features of the experiment, there are differences in the timescales involved. This is partly due to an oversimplification of the hot exciton dynamics, which is treated as a single reservoir.[23] In reality, a cascade between many different states can contribute to a richer and slower dynamics.

In summary, we have presented the full dynamics of an all-optical transistor switch, which is promising for high-speed inter-chip and intra-chip communication for core-based integrated circuits. The results are interpreted as a result of polariton propagation and energy relaxation in a dynamic potential due to the exciton reservoir, which can be optically controlled.

C.A. and G.T. acknowledge financial support from Spanish FPU and FPI scholarships, respectively. The work was partially supported by the Spanish MEC MAT2011-22997, CAM (*S*-2009/ESP-1503) and FP7 ITN's ''Clermont4'' (235114), "Spin-optronics" (237252) and INDEX (289968) projects.

*carlos.anton@uam.es